\documentclass[aps,prl,twocolumn]{revtex4-2}
\usepackage[latin9]{inputenc}
\usepackage{graphicx}
\usepackage{amssymb}
\usepackage{amsmath}
\usepackage{color, soul}
\usepackage{bm}
\usepackage[caption=false]{subfig}
\usepackage{placeins}
\usepackage{dsfont}
\usepackage{braket}

\newcommand{\w}{\omega}

\newcommand{\HMF}{\mathcal{H}_{\rm MF}}
\newcommand{\TK}{T_{\rm K}}
\newcommand{\JK}{J_{\rm K}}
\newcommand{\JH}{J_{\rm H}}
\newcommand{\nc}{n_{\rm c}}

\newcommand{\epsk}{\epsilon_{\vec{k}}}

\newcommand{\priro}{Pr$_2$Ir$_2$O$_7$}
\newcommand{\urusi}{URu$_2$Si$_2$}

\definecolor{darkgreen}{rgb}{0,0.5,0}
\definecolor{darkblue}{rgb}{0,0,0.5}
\definecolor{purple}{rgb}{0.35,0,0.35}
\definecolor{orange}{rgb}{0.9,0.4,0}

\graphicspath{{./}{./figs/}}

\usepackage{hyperref}
\hypersetup{
    colorlinks=true,
    linkcolor=blue,
    citecolor=darkblue,
    urlcolor=darkblue,
    }

\begin{document}
\title{
Emergent Chiral Metal near a Kondo Breakdown Quantum Phase Transition
}

\author{Tom Drechsler\,\href{https://orcid.org/0009-0007-2527-3514}{\includegraphics[scale=0.2]{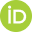}}}
\author{Matthias Vojta\,\href{https://orcid.org/0000-0002-5740-831X}{\includegraphics[scale=0.2]{ORCID-iD_icon_32x32.png}}}
\affiliation{Institut f\"ur Theoretische Physik and W\"urzburg-Dresden Cluster of Excellence ct.qmat, Technische Universit\"at Dresden,
01062 Dresden, Germany}

\date{March 13, 2025}

\begin{abstract}
The destruction of the Kondo effect in a local-moment metal can lead to a topological non-Fermi-liquid phase, dubbed fractionalized Fermi liquid, with spinon-type excitations and an emergent gauge field. We demonstrate that, if the latter displays an internal $\pi$-flux structure, a chiral heavy-fermion metal emerges near the Kondo-breakdown transition. Utilizing a parton mean-field theory describing the transition between a conventional heavy Fermi liquid and a U(1) fractionalized Fermi liquid, we find a novel intermediate phase near the transition whose emergent flux pattern spontaneously breaks both translation and time-reversal symmetries. This phase is an orbital antiferromagnet, and we derive a Landau-type theory which shows that such a phase generically emerges from a $\pi$-flux spin liquid. We discuss the relevance to pertinent experiments.
\end{abstract}

\maketitle


Many-body phases that spontaneously break time-reversal (TR) symmetry, but do not display spin dipole order, have attracted significant interest in the past and continue to do so. On the theory side, examples include insulating chiral spin liquids \cite{kalmeyer87,wen89}, TR-breaking higher-rank multipole orders \cite{kuramoto09}, and various types of orbital antiferromagnets \cite{ubbens92,varma97,morr_ddw}, many of the latter discussed in the context of doped Mott insulators. Experimentally, such phases are often considered as candidates for ``hidden order,'' as determining their character is difficult due to weak (or null) signatures in conventional scattering experiments. Multiple proposals for candidate phases were made, e.g., for the pseudogap regime of cuprate superconductors \cite{varma97,morr_ddw} as well as for the material {\urusi} \cite{hanzawa07,ikeda12}, but drawing definite conclusions has remained difficult.

In this Letter, we uncover a novel route to a TR-breaking metallic state. We utilize two bands of different correlation strength, as realized in $f$-electron metals. Such systems host a variety of fascinating phenomena, including many forms of magnetism, charge order, superconductivity, strange-metal behavior, and quantum criticality \cite{stewart01,hvl07,paschen21}. While Kondo screening of local moments leads to heavy Fermi-liquid (FL) metals, the breakdown thereof can cause nontrivial quantum phase transitions \cite{coleman01,si01,senthil03,senthil04,senthil05} and possibly topological non-Fermi-liquid states. A prominent example is that of a fractionalized Fermi liquid (FL$^\ast$), i.e., a metallic non-Kondo state encompassing local moments themselves forming a spin liquid \cite{senthil03,senthil04}. Numerous materials have been discussed as possible realizations of FL$^\ast$ phases, including $f$-electron {\priro} \cite{nakatsuji06,machida10}, CeCoIn$_5$ \cite{analytis22}, Ir-doped YbRh$_2$Si$_2$ \cite{friedemann09}, as well as underdoped cuprates \cite{moonss11,christos23}. Most recently, Moir\'e structures, e.g. of dichalcogenides, have been established as arena for heavy-fermion physics \cite{liljeroth21,zhao23}, and proposals for Kondo-breakdown phases have been made \cite{ruhman21,potter22,millis23}.

\begin{figure}[!tb]
\includegraphics[width=\columnwidth]{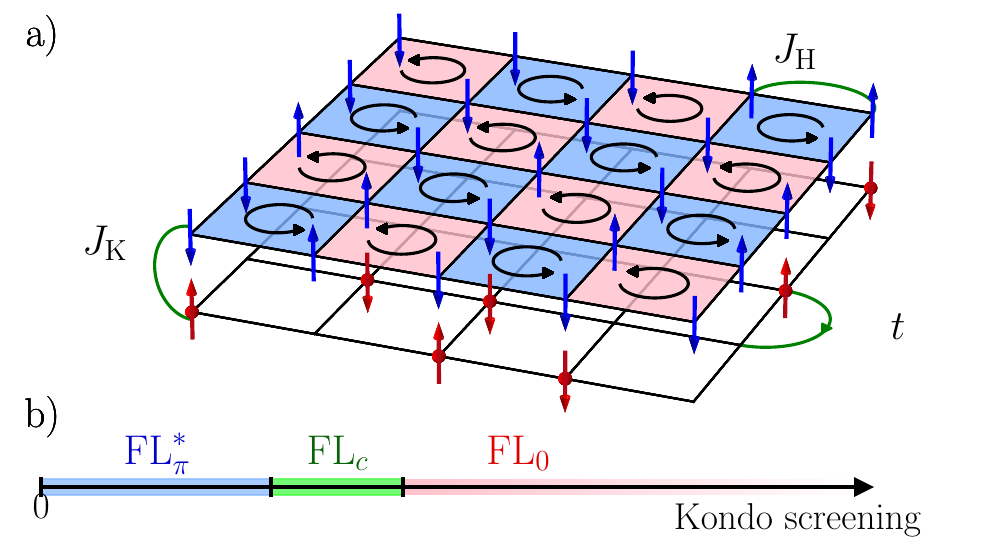}
\caption{
(a) Kondo-Heisenberg system and its intermediate chiral-metal phase, characterized by a checkerboard pattern of plaquette fluxes.
(b) Schematic zero-temperature phase diagram: The Kondo-driven transition between a fractionalized Fermi liquid with $\pi$ flux (FL$_\pi^\ast$, blue) and the conventional heavy Fermi liquid (FL$_0$, red) is generically masked by an intermediate chiral heavy-fermion phase (FL$_c$, green).
}
\label{fig:phd}
\end{figure}

The TR-breaking metal to be discussed in this Letter emerges near a FL--FL$^\ast$ quantum phase transition, the key ingredient being an emergent U(1) plaquette gauge flux of $\pi$ in the local-moment sector of the FL$^\ast$ phase (which we dub FL$^\ast_\pi$). Given that deep in the heavy FL phase, fluxes are suppressed to zero (the FL$_0$ phase), the FL--FL$^\ast$ transition must be accompanied by a rearrangement of fluxes. We discover that this happens via an intermediate heavy-fermion state in which the fluxes change continuously and hence necessarily break TR, Fig.~\ref{fig:phd}. As Kondo screening endows the local-moment electrons with a charge, this chiral Fermi-liquid phase FL$_c$ is characterized by circulating currents of heavy electrons \cite{suppl}.
We obtain explicit results using a parton mean-field theory designed to capture a U(1) FL$^\ast$ phase and its heavy-FL counterpart. For simplicity, we work on a square lattice with a $\pi$-flux Dirac spin liquid. Depending on conduction-band filling and other model parameters, the intermediate chiral metal phase can be accompanied by inhomogeneous Kondo screening. In all cases, the net flux per (enlarged) unit cell is zero, hence representing an orbital antiferromagnet. We utilize the insights from mean-field theory to derive a Landau functional for the Kondo-breakdown transition. It displays a particular nonanalytic term that turns out to be responsible for the simultaneous onset of Kondo screening and TR breaking, thus establishing the generic emergence of FL$_c$ beyond the concrete model studied.
The FL$_c$ phase discovered here may be understood as a multi-band heavy-electron version of the so-called $d$-density-wave state \cite{morr_ddw,christos23} which in fact appeared earlier as staggered flux phase \cite{ubbens92}.

Given that quantum spin liquids \cite{lee08} with emergent $\pi$ flux have been recently identified in both theory and experiment \cite{hu19,smith22,bhardwaj22}, our proposal paves an exciting way to novel metallic phases. An interesting candidate compound is {\priro}, as will be discussed below.


\paragraph{Kondo-Heisenberg model}---
We study a Kondo-Heisenberg model of conduction electrons and $f$-electron local moments in space dimension $d\geq 2$,
\begin{align}
\mathcal{H} &=
\sum_{\vec k\sigma} \epsk c^{\dagger}_{\vec k\sigma} c_{{\vec k}\sigma}
+
\JK \sum_{i\sigma\sigma'} \vec S_i \cdot c^{\dagger}_{i\sigma} \frac{\vec \tau_{\sigma\sigma'}}{2} c_{i\sigma'} \notag\\
&+
\JH \sum_{\langle ij\rangle} \vec S_i \cdot \vec S_j
\label{KH}
\end{align}
in standard notation, where $\vec\tau$ is the vector of Pauli matrices, and $\langle ij\rangle$ denote first-neighbor (and possibly second-neighbor) pairs.
The low-temperature physics of the model \eqref{KH} is governed by the competition between Kondo screening, its strength being quantified by the Kondo temperature, $\TK$ \cite{tknote}, and the exchange interaction between the local moments, $\JH$, which counteracts such screening \cite{doniach77,rkkynote}. Dominant Kondo screening, $T_K\gg J_H$, results in a paramagnetic heavy FL phase, while dominant intermoment exchange, $T_K\ll J_H$, drives the local-moment sector into either a magnetically ordered or a spin-liquid state. In the latter case, the system then realizes FL$^\ast$, i.e., a metallic spin-liquid state violating Luttinger's theorem \cite{senthil03,senthil04}. Multiple theory papers have established the existence of FL$^\ast$ phases in realistic microscopic models \cite{tsvelik16,hofmann19,danu20,chung20}. The resulting ``global'' phase diagram of heavy fermions contains magnetic transitions of Landau-Ginzburg-Wilson type as well as topological transitions characterized by the breakdown of Kondo screening and a change in the Fermi volume \cite{si_global,coleman10,mv10}.

In this Letter, we will be concerned with Kondo-breakdown transitions between FL and a particular type of FL$^\ast$ phases. The existence of FL$^\ast$ requires geometric frustration combined with strong quantum fluctuations. Since accurately solving a concrete microscopic model of the form \eqref{KH} is hard, we will employ a suitable parton theory designed to capture the formation of a spin liquid in the local-moment sector. Conceptually, a multitude of different spin liquids are possible, distinguished by (among other things) their emergent gauge symmetry. While the transition between FL and a $Z_2$ FL$^\ast$ phase is generically masked by superconductivity \cite{senthil03}, this does not hold for the U(1) case which we therefore focus on.


\paragraph{Parton mean-field theory}---
Using the Abrikosov-fermion representation of spins $\vec S_i = \sum_{\sigma\sigma'} f_{i\sigma}^{\dagger} \vec \tau_{\sigma\sigma'} f_{i\sigma'}/2$ with the constraint $\sum_\sigma f_{i\sigma}^\dagger f_{i\sigma}=1$ \cite{abrikosov65} enables a decoupling of the Kondo interaction by a bosonic field $b_i$ conjugate to $c_{i\sigma}^\dagger f_{i\sigma}$ \cite{coleman84}. The intermoment Heisenberg interaction is decoupled in the particle-hole channel by a bosonic field $\chi_{j\leftarrow i}$ conjugate to $f_{j\sigma}^\dagger f_{i\sigma}$ \cite{affleck88,marston89}, and the resulting theory enjoys an emergent U(1) gauge symmetry. A saddle-point approximation for both $b$ and $\chi$ becomes exact in a large-$N$ limit where the spin symmetry is enlarged to SU($N$). At the saddle point, nonzero $b$ and $\chi$ signal Kondo screening and spinon hopping, respectively, while SU(2) spin symmetry is generically unbroken.

Notably, in the $f$-electron sector the lowest-energy saddle point is typically spatially inhomogeneous and describes a dimerized valence-bond solid instead of a spin liquid \cite{marston89}. Within the mean-field theory, an additional biquadratic interaction can prevent the system from dimerization. Therefore we add
\begin{equation}
\mathcal{H}_{\mathrm{biq}} =
\frac{\kappa J_H}{2} \sum_{\langle ij\rangle} (\vec S_i \cdot \vec S_j)^2
\label{biquadratic_interaction}
\end{equation}
which reduces to bilinear interactions for SU(2) spins $1/2$, but has a nontrivial effect in the SU($N$)-based mean-field theory.
The mean-field Hamiltonian then becomes
\begin{align}
\HMF
&=
\sum_{k\sigma} \epsk c^{\dagger}_{k\sigma} c_{k\sigma}
-
\JK \sum_{i\sigma} ( b_i c_{i\sigma}^{\dagger} f_{i\sigma} + \mathrm{h.c.})
\notag\\
&-
\tilde{\JH} \sum_{\langle ij\rangle\sigma} \left[\chi_{j\leftarrow i}(1-2\tilde{\kappa}|\chi_{j\leftarrow i}|^2) f_{j\sigma}^{\dagger}f_{i\sigma} + \mathrm{h.c.}\right]
\notag\\
&-\sum_{i\sigma} \mu_i^f f_{i\sigma}^{\dagger} f_{i\sigma}
\label{hmf}
\end{align}
up to constants, with $\tilde{\JH} = \JH(1+\kappa/4)$, $\tilde{\kappa}=\kappa/(1+\kappa/4)$, and $\mu_i^f$ the Lagrange multiplier to enforce the occupation constraint on site $i$. The self-consistency equations can be found in the Supplemental Material \cite{suppl}.

For computational simplicity, we will obtain explicit results for a two-dimensional square lattice. Indeed, the spin-$1/2$ Heisenberg model on that lattice realizes a spin-liquid phase if a second-neighbor coupling $J_2$ is included, with $0.45 < J_2/J < 0.56$ \cite{sandvik18,becca20,liu20}. The mean-field decoupling chosen here targets such a spin liquid, and we refrain from explicitly including $J_2$. Importantly, we expect our qualitative conclusions to be more general.

\begin{figure}[!tb]
\includegraphics[width=\columnwidth]{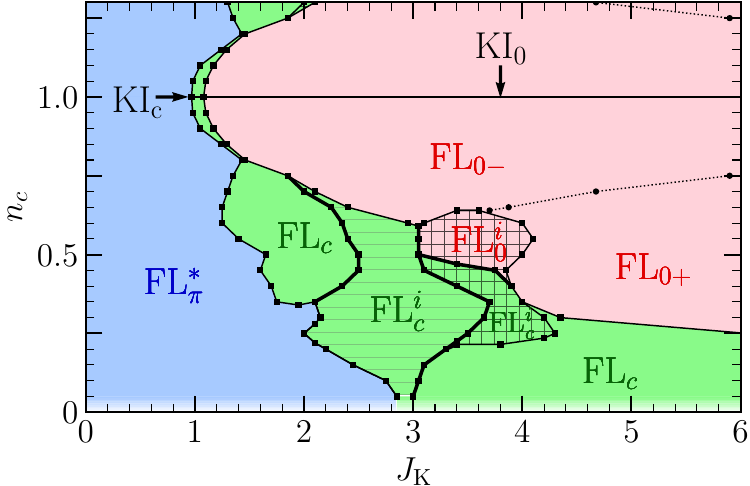}
\caption{
Low-$T$ mean-field phase diagram on the (frustrated) square lattice as function of Kondo coupling $\JK$ and conduction-band filling $\nc$. The main phases are as in Fig.~\ref{fig:phd}(b).
At half-filling, $\nc=1$, the large-$\JK$ phase is a Kondo insulator (KI); due to particle--hole symmetry the phase diagram is symmetric with respect to $\nc\leftrightarrow 2-\nc$. Thin (thick) lines denote continuous (first-order) phase transitions. Hatched phases (with upper index $i$) are inhomogeneous, i.e., display additional modulations (and may involve additional transitions which are not shown). FL$_0$ has a mean-field parameter $\chi>0$ ($\chi<0$) and is hence dubbed FL$_{0+}$ (FL$_{0-}$), respectively. Parameters are $t=J_H=1$, $\kappa=10$, $T=0.05$.
}
\label{fig:mfphd}
\end{figure}

Within the mean-field treatment, both $b_i$ and $\chi_{j\leftarrow i}$ are complex numbers. While their individual phases are not gauge-invariant, the gauge flux through each elementary plaquette, $\Phi_\square={\rm arg}(\chi_{12}\chi_{23}\chi_{34}\chi_{41})$ where $1\ldots4$ denote the sites of the plaquette, is a gauge-invariant observable and corresponds to orbital magnetism.

To account for possible breaking of spatial symmetries, we solve the mean-field equations for large real-space unit cells of $N_{sc}$ sites with periodic boundary conditions \cite{suppl}. We fix $t=1$ as the energy unit. Most results will be shown for $N_{sc}=4$, $\JH=1$ and $\kappa=10$, the latter being sufficient to stabilize a homogeneous U(1) spin liquid in the local-moment sector \cite{marston89,kappafoot}, different $\JK$, and different conduction-band fillings $\nc=(1/N_s) \sum_{i\sigma} c_{i\sigma}^\dagger c_{i\sigma}$, with $N_s$ the number of lattice sites.


\paragraph{Phase diagram}---
The low-temperature mean-field phase diagram as function of $\JK$ and $\nc$ is in Fig.~\ref{fig:mfphd}; phase diagrams at different $\nc$ as function of $\JK$ and temperature $T$ are shown in the Supplemental Material \cite{suppl}.

For small $\JK$ and low $T$ we obtain the expected FL$^\ast$ phase: here $b_i=0$ signaling the absence of Kondo screening, and the bond variables $\chi_{ij}$ display homogeneous magnitude and phases such that a $\pi$ gauge flux emerges in each $f$-electron plaquette. In this $\pi$-flux spin liquid, the spinons follow a Dirac dispersion \cite{affleck88}.
For large $\JK$, a homogeneous heavy Fermi liquid with nonzero $b_i\equiv b$ and $\chi_{ij}\equiv\chi$ is obtained, with zero plaquette fluxes. Here $b$ and $\chi$ can be chosen real, and $\chi$ is either positive or negative, which leads to phases with distinct heavy-fermion band structures, denoted as FL$_{0+}$ and FL$_{0-}$, respectively \cite{zhu08}.
The FL$^\ast$ and FL phases are separated by an intermediate phase for all $\nc$ whose width depends on both $T$ and $n_c$, which we will characterize below. At high $T$ there is an additional decoupled phase reflecting conduction electrons weakly scattering off thermally fluctuating moments.
A representative evolution of mean-field parameters as function of $\JK$ is shown in Fig.~\ref{fig:mfp1}; more results are in the Supplemental Material \cite{suppl}.

\begin{figure}[!tb]
\includegraphics[width=\columnwidth]{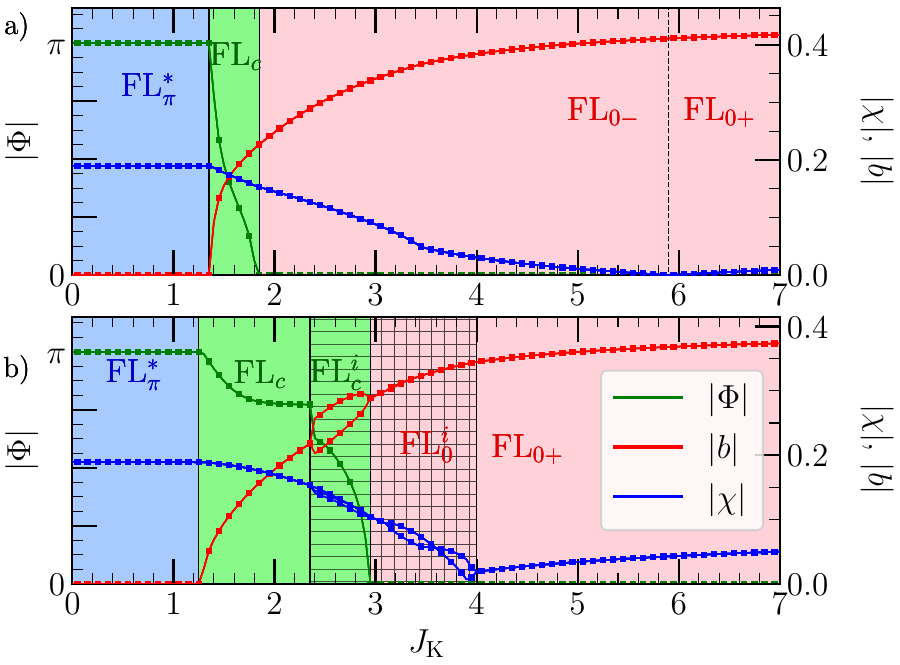}
\caption{
Evolution of the mean-field parameters $|b|$, $|\chi|$, and the plaquette fluxes $|\Phi|$ as function of $\JK$. Parameters are
(a) $\nc=0.75$,
(b) $\nc=0.6$,
and $\JH=1$, $\kappa=10$, and $T=0.05$.
Values of $\Phi$ unequal to $0,\pi$ signal spontaneous TR breaking.
While case (a) has one intermediate chiral phase bounded by two continuous transitions, case (b) displays phases (hatched) with additional modulations, for details see text.
}
\label{fig:mfp1}
\end{figure}

\begin{figure*}[!tb]
\includegraphics[width=\textwidth]{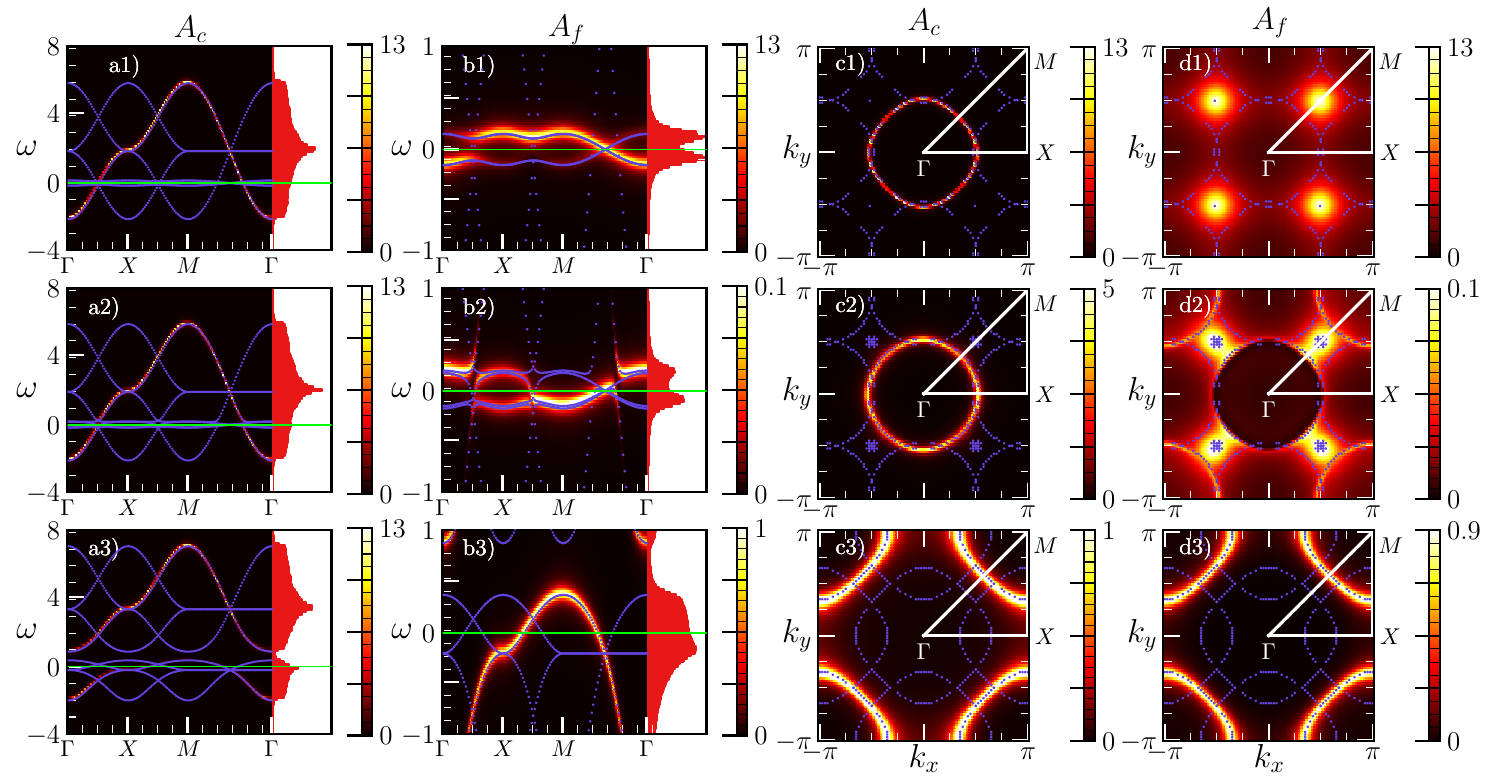}
\caption{
Single-particle spectral function of the $c$ (a,c) and $\tilde{f}$ electrons (b,d), shown as momentum-space cut (a,b) or Fermi-surface cut (c,d), for parameter values $\nc=0.4$, $T=0.05$, and $\JK = 0$ (top, FL$^\ast$), $2.0$ (middle, FL$_c$), and $4.5$ (bottom, FL$_0$). Thin lines indicate the quasiparticle bands while the color encodes the spectral intensity. In (a,b) the right insets display the momentum-integrated spectral function.
The $\tilde{f}$-electron spectrum vanishes in the FL$^\ast$ phase, therefore panels (b1,d1) show the spectrum of the auxiliary $f$ particles.
}
\label{fig:spec1}
\end{figure*}


\paragraph{Emergent chiral metal}---
The intermediate phase FL$_c$ is characterized by nonzero $b_i$, hence it is a heavy-fermion metal, and has plaquette fluxes different from 0 or $\pi$, hence breaking TR. For most parameters, we have found the solution with lowest free energy to display a four-site mean-field unit cell, with homogeneous $b_i\equiv b$ and a checkerboard arrangement of fluxes, Fig.~\ref{fig:phd}(a).
In the phase diagram, the FL$_c$ phase is most stable at low $T$ and narrows with increasing $T$. The FL$_c$ phase also narrows upon approaching $\nc=1$, but a TR-breaking flux pattern is even present at $\nc=1$ where the system transits from FL$^\ast_\pi$ to a zero-flux Kondo insulator (KI$_0$) via an intermediate chiral Kondo insulator (KI$_c$). The transitions into FL$_c$ involve the breaking of both translation and TR symmetries.
%
Within the FL$_c$ phase, we frequently observe additional transitions, as in Fig.~\ref{fig:mfp1}(b), either continuous or of first order. Some of them are accompanied by additional spatial symmetry breaking, i.e., weak modulations leading to larger unit cells, with examples shown in the Supplemental Material \cite{suppl}.


\paragraph{Single-particle spectrum}---
From the mean-field solution, we compute $A_c(\vec k,\w)$ and $A_f(\vec k,\w)$ corresponding to the spectra of $c$ and $f$ fermions in Eq.~\eqref{hmf}. We note that, while $A_c$ corresponds to the physical conduction-electron propagator, $A_f$ is that of the auxiliary fermions and hence not observable [and not invariant under U(1) gauge transformations]. We therefore also compute $A_{\tilde{f}}$ for operators $\tilde{f}^\dagger_i=b_i f^\dagger_i$ corresponding to physical $f$ electrons in an Anderson-model setting \cite{coleman84}.

Representative cuts in energy-momentum space are shown in Fig.~\ref{fig:spec1}. In the limit of small $\JK$ corresponding to FL$^\ast_\pi$, $A_c$ consists of the original light $c$ band with small Fermi surface, while $A_f$ displays Dirac points (whose position is gauge-dependent). In the opposite limit of large $\JK$ we see the hybridized heavy bands of FL$_0$, forming a large Fermi surface.
The band structure in FL$_c$ is more complicated: Its heavy-electron bands are reconstructed into multiple pockets due to the broken translation symmetry of the checkerboard flux pattern. Because of the weak Kondo screening, the $c$-electron spectral function is close to that in FL$^\ast$, Figs.~\ref{fig:spec1}(a2,c2), while the $f$-electron spectrum signals hybridization effects and a tendency to Fermi-surface enlargement.
The density-of-states plots indicate that all FL phases display a hybridization gap as expected; in contrast, the gap seen in Fig.~\ref{fig:spec1}(b1) arises from the Dirac dispersion of the spinons.


\paragraph{Landau-type theory}---
Our mean-field results show a direct and continuous transition from FL$_\pi^\ast$ to FL$_c$, which involves the onset of Kondo screening, i.e., the loss of the topological order of FL$^\ast$ \cite{senthil03,senthil04}, concomitantly with the breaking of TR and translation symmetries. We have analyzed this remarkable non-Landau behavior, present already at the mean-field level, by constructing a Landau-type theory for this transition. It involves the Kondo field $b$ and the deviation of the plaquette fluxes from $\pi$, $\tilde\Phi=\Phi-\pi$. The low-order terms read
\begin{equation}
F(b, \tilde\Phi) = c_1 \tilde\Phi^2 + c_2 \tilde\Phi^4 + d_1 |b|^2 + d_2 |b|^4 - v |b|^2 |\tilde\Phi|
\label{landau}
\end{equation}
with $c_{1,2}$, $d_2$, and $v$ being positive expansion coefficients, and $d_1$ tuning the FL$^\ast_\pi$--FL$_c$ transition. Crucially, the Dirac-spinon band structure in FL$_\pi^\ast$ is responsible for generating the last term which is nonanalytic in the gauge flux. This nonanalytic behavior implies that the onset of Kondo screening necessarily induces TR breaking, for details see Supplemental Material \cite{suppl}. A detailed study of the quantum critical behavior is left for future work.


\paragraph{Beyond mean-field; Generalizations}---
The considerations presented so far neglect fluctuation effects. We expect that the low-$T$ FL phases found in this work are stable against fluctuations --- except for the possible occurrence of superconductivity at very low $T$ --- as the condensation of $b$ gaps out the U(1) gauge field by a Higgs-type mechanism \cite{senthil04}. This can be different on the FL$^\ast$ side. In particular, precursors of Kondo screening may induce deviations of the plaquette fluxes from $\pi$ inside FL$^\ast$, leading to a fractionalized chiral FL$^\ast_c$ phase neighboring the FL$_c$ phase; similarly, higher-order RKKY interaction may also induce a chiral spin liquid in the $f$-electron sector \cite{Cookmeyer21} leading to FL$^\ast_c$, in both cases enlarging the regime of TR breaking.
In principle, the FL$^{\ast}$--FL transition could be strongly first order, omitting FL$_c$. However, we see no indications for such a scenario and conclude that TR breaking can be expected to survive beyond mean-field near the FL$^{\ast}_\pi$--FL$_0$ transition.

%
%
While our concrete results have been obtained for the square lattice, the above Landau theory shows that they qualitatively apply to any Kondo-lattice system where the local-moment sector can be brought into a topological $\pi$-flux spin liquid. Hence, a chiral metal will generically occur, Fig.~\ref{fig:phd}(b), however, its precise pattern of emergent fluxes and symmetry breaking, as well as other details of the phase diagram, will depend on the underlying lattice structure.


\paragraph{Experimental realizations}---
Two ingredients are key to the chiral metal proposed here, namely (i) a frustrated local-moment system hosting a $\pi$-flux spin liquid, and (ii) conduction electrons providing weak screening only, such that the system is proximate to a Kondo-breakdown transition.
Theoretically, a U(1) $\pi$-flux spin liquid has been deduced to exist for a number of frustrated spin models, most notably in the $J_1$-$J_2$ model on the triangular lattice \cite{hu19} and in the dipolar-octupolar spin model for quantum spin ice \cite{patri20}. The latter has been argued to be realized in Ce$_2$Zr$_2$O$_7$ \cite{smith22,bhardwaj22}, hence quantum spins on a pyrochlore lattice appear promising for (i) \cite{benton18}.

To also satisfy (ii), we suggest {\priro} as a prime candidate material.
Its local moments display icelike correlations, and it shows no magnetic long-range order down to lowest temperatures \cite{nakatsuji06}. Its carrier concentration is small due to the proximity to a quadratic band-touching point. A regime of divergent magnetic Gr\"uneisen parameter \cite{pr2ir2o7_grueneisen} indicates the proximity of the material to a QPT.
Remarkably, {\priro} displays an anomalous Hall effect below 6\,K, indicative of a TR-breaking state without magnetic dipole order; freezing of magnetic moments sets in at a much lower temperature of 2\,K \cite{machida10}. We therefore conjecture that {\priro} realizes a variant of the chiral metal proposed here.
We note that STM experiments \cite{kavai21} indicate that {\priro} crystals are subject to substantial quenched disorder. Such disorder will likely lead to a pinning of the flux pattern and therefore broaden the regime of the chiral metal. Theoretical investigations along these lines are underway.


\paragraph{Summary}---
We have shown that a chiral heavy-fermion metal emerges naturally near a Kondo-breakdown quantum phase transition, provided that the proximate FL$^\ast$ phase is of U(1) $\pi$-flux type. We have characterized this phase in terms of mean-field order parameters and electronic spectral functions. We propose that the pyrochlore compound {\priro} hosts a realization of this physics, given that it displays a TR-breaking phase without dipole order at low temperatures. We suggest future low-temperature electronic structure measurements to uncover the spatial symmetry breaking proposed here. Similarly, searching for other pyrochlore local-moment metals with weak Kondo screening is a promising route to the said chiral metal.


\paragraph{Note added}---
Recently, we became of aware of Ref.~\onlinecite{koenig23} where a different route to a TR-breaking metallic phase near a Kondo-breakdown transition was suggested.

\acknowledgments

We thank F. F. Assaad, P. M. C\^onsoli, B. Danu, B. Frank, T. Grover, L. Janssen, E. J. K\"onig, and Z. H. Liu for discussions and collaborations on related work.
Financial support from the Deutsche Forschungsgemeinschaft through SFB 1143 (Project-id 247310070) and the W\"urzburg-Dresden Cluster of Excellence on Complexity and Topology in Quantum Matter -- \textit{ct.qmat} (EXC 2147, Project-id 390858490) is gratefully acknowledged.



\begin{thebibliography}{99}

\bibitem{kalmeyer87}
V. Kalmeyer and R. B. Laughlin,
\href{https://doi.org/10.1103/PhysRevLett.59.2095}{Phys. Rev. Lett. \textbf{59}, 2095 (1987)}.

\bibitem{wen89}
X. G. Wen, F. Wilczek, and A. Zee,
\href{https://doi.org/10.1103/PhysRevB.39.11413}{Phys. Rev. B \textbf{39}, 11413 (1989)}.

\bibitem{kuramoto09}
Y. Kuramoto, H. Kusunose, and A. Kiss,
\href{https://doi.org/10.1143/JPSJ.78.072001}{J. Phys. Soc. Jpn. \textbf{78}, 072001 (2009)}.

\bibitem{ubbens92}
M. U. Ubbens and P. A. Lee,
\href{https://doi.org/10.1103/PhysRevB.46.8434}{Phys. Rev. B \textbf{46}, 8434 (1992)}.

\bibitem{varma97}
C. M. Varma,
\href{https://doi.org/10.1103/PhysRevB.55.14554}{Phys. Rev. B \textbf{55}, 14554 (1997)}.

\bibitem{morr_ddw}
S. Chakravarty, R. B. Laughlin, D. K. Morr, and C. Nayak,
\href{https://doi.org/10.1103/PhysRevB.63.094503}{Phys. Rev. B \textbf{63}, 094503 (2001)}.

\bibitem{hanzawa07}
K. Hanzawa,
\href{https://doi.org/10.1088/0953-8984/19/7/072202}{J. Phys.: Condens. Matter \textbf{19}, 072202 (2007)}.

\bibitem{ikeda12}
H. Ikeda, M.-T. Suzuki, R. Arita, T. Takimoto, T. Shibauchi, and Y. Matsuda,
\href{https://doi.org/10.1038/nphys2330}{Nat. Phys. \textbf{8}, 528 (2012)}.

\bibitem{stewart01}
G. R. Stewart,
\href{https://doi.org/10.1103/RevModPhys.73.797}{Rev. Mod. Phys. {\bf 73}, 797 (2001)}.

\bibitem{hvl07}
H. v. L\"ohneysen, A. Rosch, M. Vojta, and P. W\"olfle,
\href{https://doi.org/10.1103/RevModPhys.79.1015}{Rev. Mod. Phys. {\bf 79}, 1015 (2007)}.

\bibitem{paschen21}
S. Paschen and Q. Si,
\href{https://doi.org/10.1038/s42254-020-00262-6}{Nat. Rev. Phys. \textbf{3}, 9 (2021)}.


\bibitem{coleman01}
P. Coleman, C. Pepin, Q. Si, and R. Ramazashvili,
\href{https://doi.org/10.1088/0953-8984/13/35/202}{J. Phys. Condens. Matter \textbf{13}, R723 (2001)}.

\bibitem{si01}
Q. Si, S. Rabello, K. Ingersent, and J. L. Smith,
\href{https://doi.org/10.1038/35101507}{Nature (London) \textbf{413}, 804 (2001)}.

\bibitem{senthil03}
T. Senthil, S. Sachdev, and M. Vojta,
\href{https://doi.org/10.1103/PhysRevLett.90.216403}{Phys. Rev. Lett. \textbf{90}, 216403 (2003)}.

\bibitem{senthil04}
T. Senthil, M. Vojta, and S. Sachdev,
\href{https://doi.org/10.1103/PhysRevB.69.035111}{Phys. Rev. B \textbf{69}, 035111 (2004)}.

\bibitem{senthil05}
\href{https://doi.org/10.1016/j.physb.2004.12.041}{T. Senthil, S. Sachdev, and M. Vojta, Physica B (Amsterdam) \textbf{359-361}, 9 (2005)}.

\bibitem{nakatsuji06}
S. Nakatsuji, Y. Machida, Y. Maeno, T. Tayama, T. Sakakibara, J. van Duijn, L. Balicas, J. N. Millican, R. T. Macaluso, and J. Y. Chan,
\href{https://doi.org/10.1103/PhysRevLett.96.087204}{Phys. Rev. Lett. \textbf{96}, 087204 (2006)}.

\bibitem{machida10}
Y. Machida, S. Nakatsuji, S. Onoda, T. Tayama, and T. Sakakibara,
\href{https://doi.org/10.1038/nature08680}{Nature (London) \textbf{463}, 210 (2010)}.

\bibitem{analytis22}
N. Maksimovic {\em et al.},
\href{https://doi.org/10.1126/science.aaz4566}{Science \textbf{375}, 76 (2022)}.

\bibitem{friedemann09}
S. Friedemann, T. Westerkamp, M. Brando, N. Oeschler, S. Wirth, P. Gegenwart, C. Krellner, C. Geibel, and F. Steglich,
\href{https://doi.org/10.1038/nphys1299}{Nat. Phys. \textbf{5}, 465 (2009)}.

\bibitem{moonss11}
E. G. Moon and S. Sachdev,
\href{https://doi.org/10.1103/PhysRevB.83.224508}{Phys. Rev. B \textbf{83}, 224508 (2011)}.

\bibitem{christos23}
M. Christos, Z.-X. Luo, H. Shackleton, Y.-H. Zhang, M. Scheurer, and S. Sachdev,
\href{https://doi.org/10.1073/pnas.2302701120}{Proc. Natl. Acad. Sci. U.S.A \textbf{120}, e2302701120 (2023)}.

\bibitem{liljeroth21}
V. Vano, M. Amini, S. C. Ganguli, G. Chen, J. L. Lado, S. Kezilebieke, and P. Liljeroth,
\href{https://doi.org/10.1038/s41586-021-04021-0}{Nature (London) \textbf{599}, 582 (2021)}.

\bibitem{zhao23}
W. Zhao, B. Shen, Z. Tao, Z. Han, K. Kang, K. Watanabe, T. Taniguchi, K. F. Mak, and J. Shan,
\href{https://doi.org/10.1038/s41586-023-05800-7}{Nature (London) \textbf{616}, 61 (2023)}.

\bibitem{ruhman21}
A. Dalal and J. Ruhman,
\href{https://doi.org/10.1103/PhysRevResearch.3.043173}{Phys. Rev. Res. 3, 043173 (2021).}

\bibitem{potter22}
A. Kumar, N. C. Hu, A. H. MacDonald, and A. C. Potter,
\href{https://doi.org/10.1103/PhysRevB.106.L041116}{Phys. Rev. B \textbf{106}, L041116 (2022)}.

\bibitem{millis23}
D. Guerci, J. Wang, J. Zang, J. Cano, J. H. Pixley, and A. Millis,
\href{https://doi.org/10.1126/sciadv.ade7701}{Sci. Adv. \textbf{9}, eade7701 (2023)}.

\bibitem{suppl}
See Supplemental Material at \url{https://doi.org/10.1103/PhysRevLett.134.106503}, which includes Ref.~\onlinecite{burdin00}, for details of the mean-field theory, additional numerical results, and an analysis of the non-Landau character of the FL$^\ast_\pi$--FL$_c$ transition.

\bibitem{burdin00}
S.~Burdin, A.~Georges, and D.~R.~Grempel,
\href{https://doi.org/10.1103/PhysRevLett.85.1048}{Phys. Rev. Lett. \textbf{85}, 1048 (2000)}.

\bibitem{lee08}
P. A. Lee,
\href{https://doi.org/10.1126/science.1163196}{Science \textbf{321}, 1306 (2008)}.

\bibitem{hu19}
S. Hu, W. Zhu, S. Eggert, and Y.-C. He,
\href{https://doi.org/10.1103/PhysRevLett.123.207203}{Phys. Rev. Lett. \textbf{123}, 207203 (2019)}.

\bibitem{smith22}
E. M. Smith \textit{et al.}, 
\href{https://doi.org/10.1103/PhysRevX.12.021015}{Phys. Rev. X \textbf{12}, 021015 (2022)}.

\bibitem{bhardwaj22}
A. Bhardwaj, S. Zhang, H. Yan, R. Moessner, A. H. Nevidomskyy, and H. J. Changlani,
\href{https://doi.org/10.1038/s41535-022-00458-2}{npj Quantum Mater. \textbf{7}, 51 (2022)}.


\bibitem{doniach77}
S. Doniach, \href{https://doi.org/10.1016/0378-4363(77)90190-5}{Physica B+C (Amsterdam) \textbf{91}, 231 (1977)}.

\bibitem{tknote}
The Kondo temperature $\TK$ is an emergent scale below which local moments get screened in the absence of intermoment coupling; we assume the standard dependence $\TK \sim D \exp(-D/\JK)$ where $D$ is the conduction-electron bandwidth of the dispersion $\epsk$ \cite{doniach77}.

\bibitem{rkkynote}
The local moments in \eqref{KH} also interact via the indirect RKKY interaction; in our qualitative discussion we assume its effect to be absorbed in $\JH$.

\bibitem{tsvelik16}
A. M. Tsvelik,
\href{https://doi.org/10.1103/PhysRevB.94.165114}{Phys. Rev. B \textbf{94}, 165114 (2016)}.

\bibitem{hofmann19}
J. S. Hofmann, F. F. Assaad, and T. Grover,
\href{https://doi.org/10.1103/PhysRevB.100.035118}{Phys. Rev. B \textbf{100}, 035118 (2019)}.

\bibitem{danu20}
B. Danu, M. Vojta, F. F. Assaad, and T. Grover,
\href{https://doi.org/10.1103/PhysRevLett.125.206602}{Phys. Rev. Lett. \textbf{125}, 206602 (2020)}.

\bibitem{chung20}
J. Wang, Y.-Y. Chang, C.-Y. Mou, S. Kirchner, and C.-H. Chung,
\href{https://doi.org/10.1103/PhysRevB.102.115133}{Phys. Rev. B \textbf{102}, 115133 (2020)}.

\bibitem{si_global}
Q. Si, \href{https://doi.org/10.1016/j.physb.2006.01.156}{Physica B (Amsterdam) \textbf{378-380}, 23 (2006)};
\href{https://doi.org/10.1002/pssb.200983082}{Phys. Status Solidi B \textbf{247}, 476 (2010)}.

\bibitem{coleman10}
P. Coleman,
\href{https://doi.org/10.1007/s10909-010-0213-4}{J. Low Temp. Phys. \textbf{161}, 182 (2010)}.

\bibitem{mv10}
M. Vojta,
\href{https://doi.org/10.1007/s10909-010-0206-3}{J. Low Temp. Phys. \textbf{161}, 203 (2010)}.

\bibitem{abrikosov65}
A. A. Abrikosov,
\href{https://doi.org/10.1103/PhysicsPhysiqueFizika.2.5}{Phys. Phys. Fiz. \textbf{2}, 5 (1965)}.

\bibitem{coleman84}
P. Coleman,
\href{https://doi.org/10.1103/PhysRevB.29.3035}{Phys. Rev. B \textbf{29}, 3035 (1984)}.

\bibitem{affleck88}
I. Affleck and J. B. Marston,
\href{https://doi.org/10.1103/PhysRevB.37.3774}{Phys. Rev. B \textbf{37}, 3774 (1988)}.

\bibitem{marston89}
J. B. Marston and I. Affleck,
\href{https://doi.org/10.1103/PhysRevB.39.11538}{Phys. Rev. B \textbf{39}, 11538 (1989)}.

\bibitem{sandvik18}
L. Wang and A. W. Sandvik,
\href{https://doi.org/10.1103/PhysRevLett.121.107202}{Phys. Rev. Lett. \textbf{121}, 107202 (2018)}.

\bibitem{becca20}
F. Ferrari and F. Becca,
\href{https://doi.org/10.1103/PhysRevB.102.014417}{Phys. Rev. B \textbf{102}, 014417 (2020)}.

\bibitem{liu20}
W.-Y. Liu, S.-S. Gong, Y.-B. Li, D. Poilblanc, W.-Q. Chen, and Z.-C. Gu,
\href{https://doi.org/10.1016/j.scib.2022.03.010}{Sci. Bull. \textbf{67}, 1034 (2022)}. 



\bibitem{kappafoot}
In the mean-field theory for the square-lattice antiferromagnet, inhomogeneous solutions such as valence-bond solids are suppressed for any $\kappa>1.9$ \cite{marston89}. With the Kondo coupling included, we find that a significant tendency towards inhomogeneous FL solutions for $\kappa\lesssim 10$, see Ref.~\onlinecite{suppl} for a more detailed discussion.

\bibitem{zhu08}
J.-X. Zhu, I. Martin, and A. R. Bishop,
\href{https://doi.org/10.1103/PhysRevLett.100.236403}{Phys. Rev. Lett. \textbf{100}, 236403 (2008)}.

\bibitem{Cookmeyer21}
T. Cookmeyer, J. Motruk, and J. E. Moore,
\href{https://doi.org/10.1103/PhysRevLett.127.087201}{Phys. Rev. Lett. \textbf{127}, 087201 (2021)}.

\bibitem{patri20}
A. S. Patri, M. Hosoi, and Y. B. Kim,
\href{https://doi.org/10.1103/PhysRevResearch.2.023253}{Phys. Rev. Res. \textbf{2}, 023253 (2020)}.

\bibitem{benton18}
O. Benton, L. D. C. Jaubert, R. R. P. Singh, J. Oitmaa, and N. Shannon,
\href{https://doi.org/10.1103/PhysRevLett.121.067201}{Phys. Rev. Lett. \textbf{121}, 067201 (2018)}.

\bibitem{pr2ir2o7_grueneisen}
Y. Tokiwa, J. J. Ishikawa, S. Nakatsuji, and P. Gegenwart,
\href{https://doi.org/10.1038/nmat3900}{Nat. Mater. {\bf 13}, 356 (2014)}.

\bibitem{kavai21}
M. Kavai \textit{et al.},
\href{https://doi.org/10.1038/s41467-021-21698-z}{Nat. Commun. \textbf{12}, 1377 (2021)}.

\bibitem{koenig23}
E. J. K\"onig,
\href{https://doi.org/10.1103/PhysRevResearch.6.L012058}{Phys. Rev. Res. \textbf{6}, L012058 (2024)}.



\end{thebibliography}
\end{document}


\title{
Supplemental material for: \\
Emergent Chiral Metal near a Kondo Breakdown Quantum Phase Transition
}

\author{Tom Drechsler}
\author{Matthias Vojta}
\affiliation{Institut f\"ur Theoretische Physik and W\"urzburg-Dresden Cluster of Excellence ct.qmat, Technische Universit\"at Dresden,
01062 Dresden, Germany}

\date{March 13, 2025}

\maketitle


\section{Parton mean-field theory}

Here we present details of the parton mean-field theory described in the main text. Using the Abrikosov-fermion representation and decoupling both the Kondo and Heisenberg interactions, the latter with its biquadratic counterpart \cite{marston89}, in the particle-hole channel yields a real-space mean-field theory of the following form for the grand-canonical operator $\mathcal{K}=\mathcal{H}-\mu^c N_c$ with $\mu^c$ the conduction-electron chemical potential:
\begin{align}
\KMF =
&- t \sum_{\langle ij\rangle\sigma} (c^{\dagger}_{i\sigma} c_{j\sigma} + \mathrm{h.c.})
-
\JK \sum_{i\sigma} (b_i c_{i\sigma}^{\dagger} f_{i\sigma} + \mathrm{h.c.})
\notag\\
&-
\tilde{\JH} \sum_{\langle ij\rangle\sigma} \left[ \chi_{j\leftarrow i}(1-2\tilde{\kappa}|\chi_{j\leftarrow i}|^2) f_{j\sigma}^{\dagger}f_{i\sigma} + \mathrm{h.c.}\right]
\notag\\
&-\sum_{i\sigma} \left[\mu^c c^{\dagger}_{i\sigma} c_{i\sigma} + \mu_i^f (f_{i\sigma}^{\dagger} f_{i\sigma} - 1) \right]
\notag\\
&+
2 \tilde{\JH} \sum_{\langle ij\rangle} |\chi_{j\leftarrow i}|^2 (1 - 3\tilde{\kappa} |\chi_{j\leftarrow i}|^2 )
\notag \\
&+
2 \JK \sum_i |b_i|^2.
\label{Mean_Field_Operator}
\end{align}
Here we have assumed nearest-neighbor hopping for the $c$-electron kinetic energy, $\epsilon_{\vec k} = -2t (\cos k_x + \cos k_y)$, on the square lattice. Further we recall that $\tilde{\JH} = \JH(1+\kappa/4)$ and $\tilde{\kappa}=\kappa/(1+\kappa/4)$.
The relevant self-consistency equations read:
\begin{align}
\chi_{j\leftarrow i} &=
\frac{1}{2}\sum_{\sigma} \langle f_{i\sigma}^{\dagger} f_{j\sigma}\rangle,
\label{Mean_Field_chi}
\\
b_{i} &=
\frac{1}{2}\sum_{\sigma} \langle f_{i\sigma}^{\dagger} c_{i\sigma}\rangle,
\label{Mean_Field_b}
\\
1 &=
\sum_{\sigma} \langle f_{i\sigma}^{\dagger} f_{i\sigma}\rangle,
\label{Mean_Field_mu_f}
\\
n_c N_s &=
\sum_{i\sigma} \langle c_{i\sigma}^{\dagger} c_{i\sigma}\rangle,
\label{Mean_Field_mu_f}
\end{align}
with $N_s$ the number of lattice sites and $\nc\in[0,2]$ the conduction-band filling.

For fixed values of the mean-field parameters, $\KMF$ is diagonalized for a real-space unit cell of $N_{sc}$ sites, yielding a matrix of size $2N_{sc} \times 2N_{sc}$. We further assume a periodic arrangement of $N_c$ such cells, enabling super-cell Fourier transformation, to improve energy resolution, such that $N_s = N_{sc} N_c$. In practice, we have used $N_{sc} = 2\times 2, N_c=6\times 6$ and $N_{sc}=4\times 4, N_c=3\times 3$. Along selected cuts through parameter space, we have used larger values of $N_c$ and determined the locations of the phase boundaries in the thermodynamic limit by finite-size extrapolation; those deviate from the ones quoted in the paper by at most 5\%.

Using the diagonalized form of $\KMF$, the equations \eqref{Mean_Field_chi}--\eqref{Mean_Field_mu_f} are solved iteratively at finite fixed temperature, using random mixing to improve convergence, starting from a random initialization of the mean-field parameters. We have used up to 50 different initializations per parameter set and eventually selected the saddle point with the lowest free energy
\begin{align}
F =
& -T \sum_{i=1}^{2N_s} \ln\lbrace 1 + \mathrm{e}^{-\beta E_i}\rbrace
+ \mu^c N_s + \sum_i \mu^f_i
\label{Free_Energy} \\
& + 2 \tilde{\JH} \sum_{\langle ij\rangle} |\chi_{j\leftarrow i}|^2 (1 - 3\tilde{\kappa} |\chi_{j\leftarrow i}|^2 )
+ 2 \JK \sum_i |b_i|^2
\notag
\end{align}
where $E_i$ are the single-particle eigenvalues of $\KMF$ and $\beta$ is the inverse temperature.

\begin{figure}[!tb]
\includegraphics[width=\columnwidth]{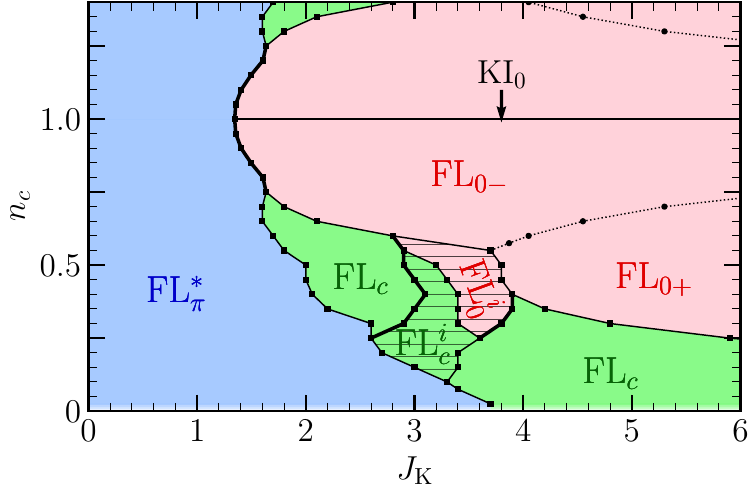}
\caption{
Mean-field phase diagram as in Fig.~1(b) of the main text, but for a higher temperature of $T=0.1$. Thin (thick) lines denote continuous (first-order) phase transitions; hatched regions correspond to phases with additional inhomogeneities.
}
\label{fig:phd2}
\end{figure}

The spectral-function plots were generated with a momentum resolution corresponding to $N_s=100\times100$. The discrete $\delta$ peaks were broadened using a Lorentzian of width $\eta=0.05$ (in units of the hopping $t$). The thin lines in Figs.~3(a,b) and \ref{fig:spec2}(a,b) simply indicate the single-particle energies, while those in Figs.~3(c,d) and \ref{fig:spec2}(c,d) indicate those momenta where a single-particle energy falls within an energy window centered at the Fermi level, with a suitably chosen window width (Fig~3: 0.02, 0.01, 0.0075 from top to bottom).


\begin{figure}[!tb]
\includegraphics[width=\columnwidth]{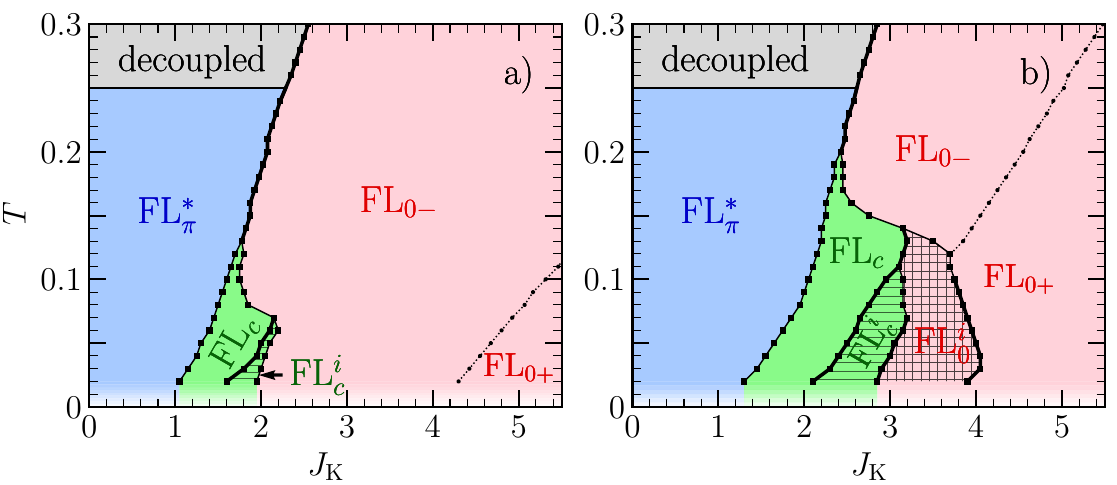}
\caption{
Mean-field phase diagrams as function of temperature $T$ and Kondo couplings $\JK$ for
(a) $\nc=0.7$ and (b) $\nc=0.5$;
other parameter values are $t=\tilde{J}_{\mathrm{H}}=1$ and $\tilde{\kappa}=10$.
}
\label{fig:phdTJK1}
\end{figure}

\section{Additional numerical results}

The following mean-field results complement those given in the main text.


\subsection{Phase diagram and temperature dependence}

We start with a phase diagram similar to Fig.~1(b) of the main paper, but for a higher temperature. Fig.~\ref{fig:phd2} shows that the chiral intermediate phase continues to be present, but the additional inhomogeneities tend to be suppressed. This can be seen in more detail in the phase diagrams as function of temperature $T$ and Kondo coupling $\JK$, obtained for $t=1$, $\tilde{J}_{\mathrm{H}}=1$ and $\tilde{\kappa}=10$ for different conduction band fillings $\nc$, shown in Fig.~\ref{fig:phdTJK1}. A decoupled phase with $b=\chi=0$ emerges at elevated $T$; its boundary to FL$^\ast$ is located at $T=\tilde{J}_{\mathrm{H}}/4$. Beyond mean-field, this elevated-temperature regime corresponds to conduction electrons weakly scattering off thermally fluctuating local moments, and the mean-field transitions FL$\leftrightarrow$decoupled and FL$^\ast${}$\leftrightarrow$decoupled get smeared by gauge fluctuations, i.e., become crossovers \cite{senthil03}.

As noted in the main text, the intermediate chiral phase FL$_c$ is most extended at low $T$ for all $\nc$. Similarly, additional inhomogeneous phases (FL$_c^i$, FL$_0^i$) get suppressed with increasing temperature. Furthermore we see from Figs. 1b, \ref{fig:phd2}, and \ref{fig:phdTJK1} that decreasing $\nc$ reduces the propensity for Kondo screening, such that the transitions FL$^\ast_\pi\leftrightarrow$FL$_c\leftrightarrow$FL$_0$ are shifted to larger values of $\JK$. Given the results of Ref.~\onlinecite{burdin00}, we expect that these transitions shift to $\JK\to\infty$ for $\nc\to 0$.

We have also computed phase diagrams for different values of $\tilde{J}_{\mathrm{H}}$, with results qualitatively similar to the ones shown here for $\tilde{J}_{\mathrm{H}}=1$. Lowering $\tilde{J}_{\mathrm{H}}$ while keeping all other parameters fixed reduces the extend of the FL$^\ast$ phase.


\begin{figure}[!t]
\includegraphics[width=\columnwidth]{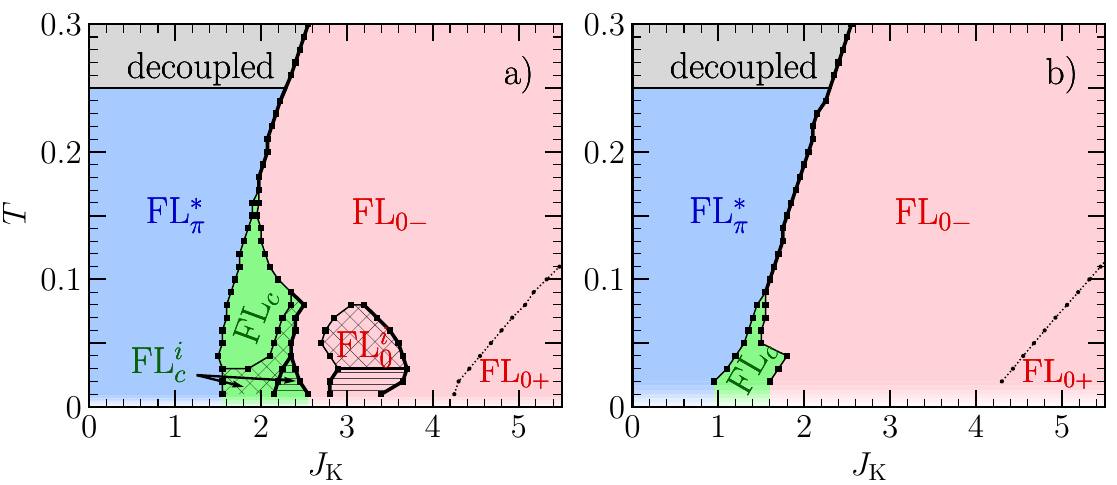}
\caption{
As in Fig.~\ref{fig:phdTJK1}, but now for $n_c=0.7$ and (a) $\tilde{\kappa}=5$ and (b) $\tilde{\kappa}=20$; for comparison the corresponding result for $\tilde{\kappa}=10$ is in Fig.~\ref{fig:phdTJK1}(a).
}
\label{fig:phdTJK2}
\end{figure}

\subsection{Influence of biquadratic coupling $\kappa$}

For comparison we also show phase diagrams for different strengths $\tilde{\kappa}$ of the biquadratic coupling in Fig.~\ref{fig:phdTJK2}. Three trends are obvious: (i) Larger $\tilde{\kappa}$ tend to favor FL over FL$^\ast$, thus shifting all transitions to smaller $\JK$. This is related to the energetic cost of the biquadratic term which is larger in FL$^\ast$ (compared to FL) because $\chi$ is larger there. (ii) As noted in the main text, larger $\tilde{\kappa}$ tends to suppress modulations of $\chi$, in line with the early results of Ref.~\onlinecite{marston89}. (iii) Larger $\tilde{\kappa}$ also tends to suppress the chiral intermediate phase FL$_c$, essentially again because it disfavors large $\chi$.

\begin{figure}[!b]
\includegraphics[width=\columnwidth]{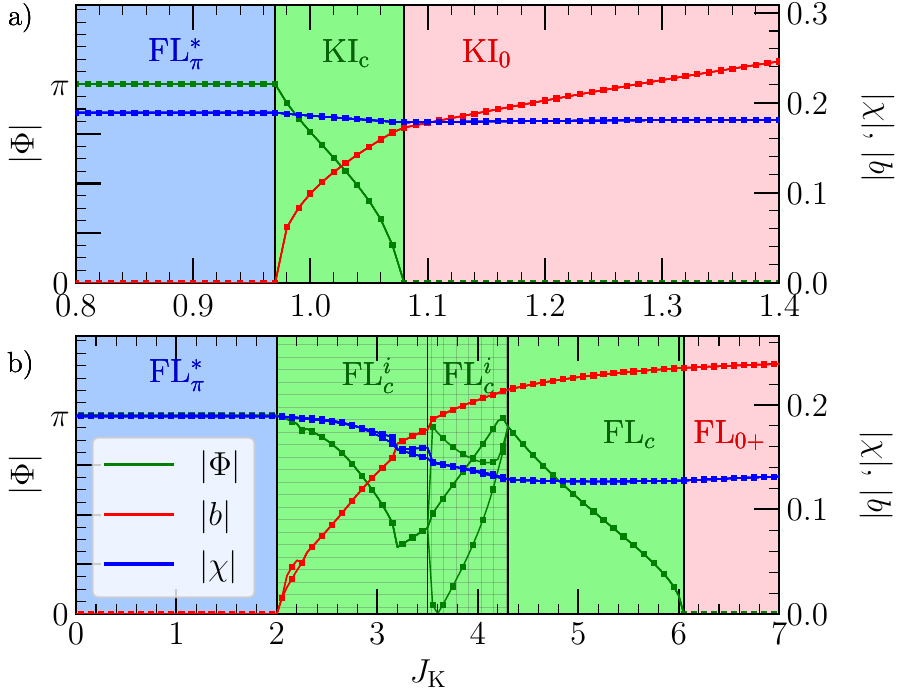}
\caption{
Evolution of the mean-field parameters as in Fig.~2 of the main text, but now for (a) $\nc=1$ and (b) $\nc=0.25$. Panel (a) illustrates the emergence of a chiral Kondo insulator at half filling, while panel (b) shows two FL$_c^i$ phases, the first one having a striped structure of $b$ and $\chi$ parameters, while the second one has an inhomogeneous distribution of fluxes $|\Phi|$ in addition, see also Fig.~\ref{fig:patt1}.
}
\label{fig:mfp2}
\end{figure}

We recall that the biquadratic coupling was introduced \cite{marston89} to stabilize a homogeneous spin-liquid solution in the local-moment sector of FL$^\ast$, i.e., to cope with the dichotomy that exact numerics for the $J_1$-$J_2$ model proves the existence of homogeneous spin liquids \cite{sandvik18,becca20,liu20} but the ``bare'' mean-field has a strong tendency towards dimerization. We therefore consider (i) and (iii) as side effects of large $\tilde{\kappa}$, not present a solution of a $J_1$-$J_2$ model beyond mean field.

We also note that, for the pure local-moment model, values of $\tilde{\kappa}>1.9$ are sufficient to suppress inhomogeneous solutions; smaller values of $\tilde{\kappa}$ lead to box or stripe modulations \cite{marston89}. For the Kondo-Heisenberg model we have found that such modulations continue to appear in the FL phase even for larger values of $\tilde{\kappa}$, see Figs. 1(b), \ref{fig:phd2}, \ref{fig:phdTJK1}, \ref{fig:phdTJK2}. The value of $\tilde{\kappa}=10$ used in the main text can therefore be seen as a compromise.


\begin{figure}[!tb]
\includegraphics[width=\columnwidth]{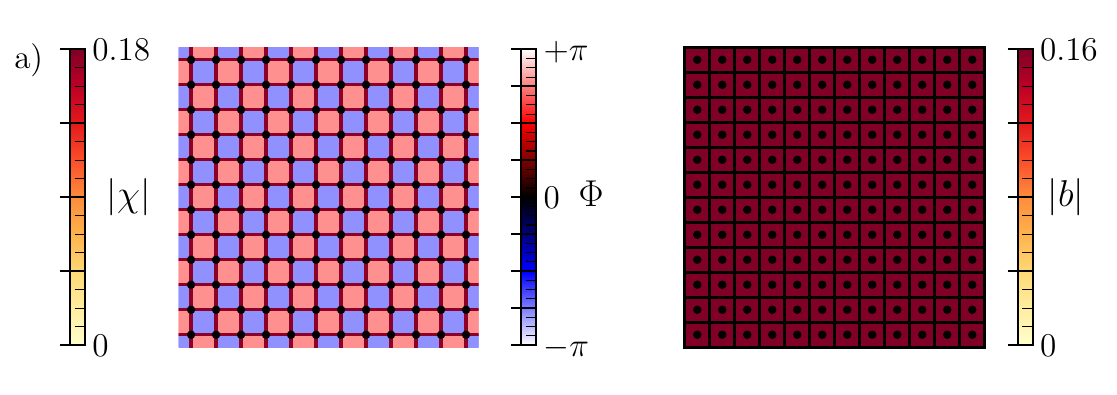}
\includegraphics[width=\columnwidth]{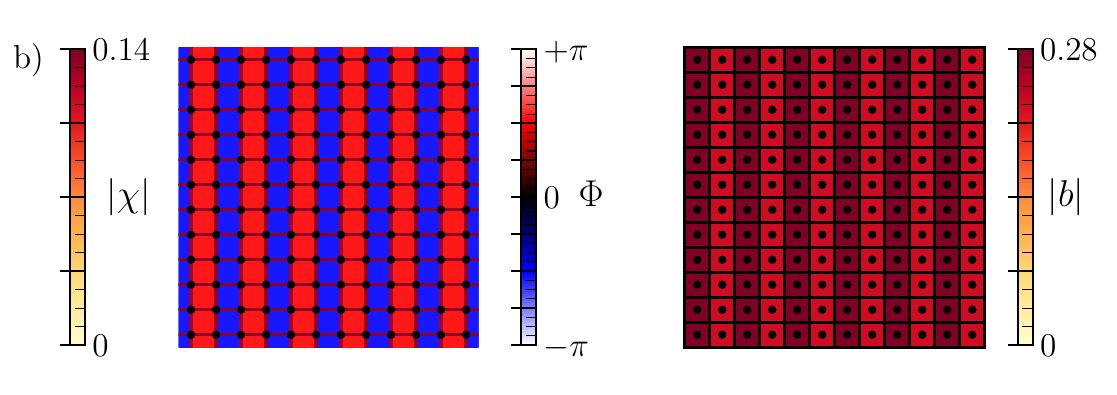}
\includegraphics[width=\columnwidth]{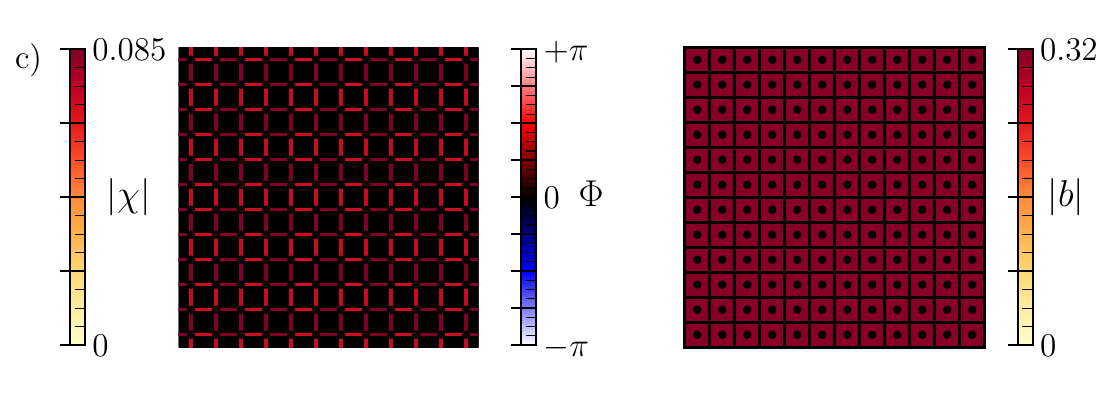}
\caption{
Illustration of the spatial distribution of the mean-field parameters, obtained from calculations using a unit cell with $N_{sc}=2\times2$ sites. Left panels show the bond variables $|\chi|$ and the plaquette fluxes $\Phi$, while the right panels show the on-site Kondo parameters $|b|$. All results are for $\tilde{J}_{\mathrm{H}}=t=1$, $\tilde{\kappa}=10$, $T=0.05$, and $\nc=0.6$.
(a) $\JK=1.9$: checkerboard flux pattern with homogeneous $|\chi|$ and $|b|$, as mostly realized in the FL$_c$ phase.
(b) $\JK=2.6$: stripe modulation of $|b|$ and $\Phi$, as an example for a FL$_c^i$ phase.
(c) $\JK=3.3$: box modulation of $|\chi|$ with $\Phi=0$, as an example for a FL$_0^i$ phase.
}
\label{fig:patt1}
\end{figure}

\begin{figure}[!tb]
\includegraphics[width=\columnwidth]{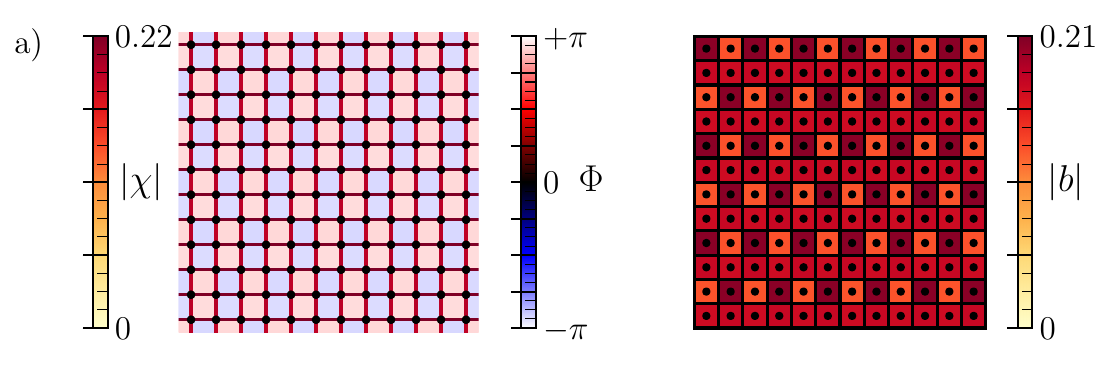}
\includegraphics[width=\columnwidth]{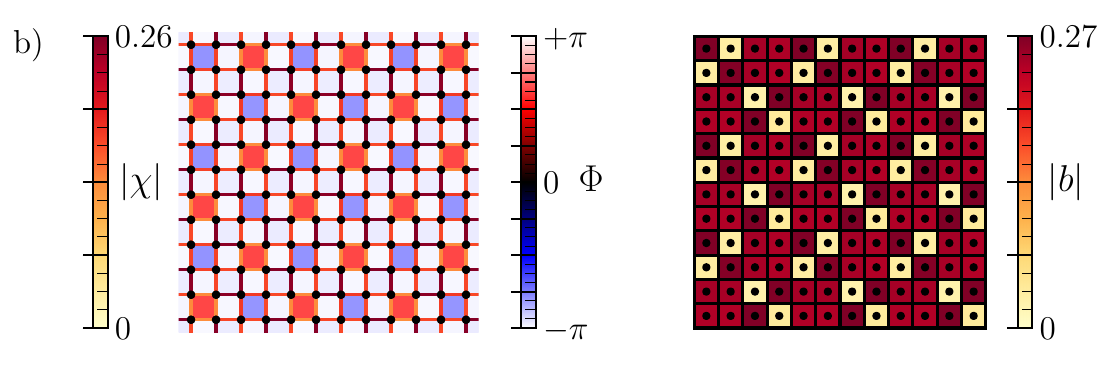}
\caption{
As in Fig.~\ref{fig:patt1}, but obtained for a larger mean-field unit cell of $N_{sc}=4\times4$ and for selected parameter values where the mean-field solution adopts this larger unit cell.
(a) $\nc=0.6$ and $\JK=2.4$.
(b) $\nc=0.5$ and $\JK=3.11$.
Both cases correspond to a particular realization of FL$_c^i$. We note that the net flux per unit cell in case (b) remains zero, as in the other solutions.
}
\label{fig:patt2}
\end{figure}

\subsection{Modulation of mean-field parameters}

Additional results for the evolution of mean-field parameters are in Fig.~\ref{fig:mfp2}. Panel (a) illustrates the behavior at half-filling where a chiral intermediate phase intervenes as well; this is a Kondo insulator, i.e., the band gap opens concomitantly with the onset of Kondo screening. Panel (b) is another case where, in addition to an FL$_c$ phase with checkerboard flux pattern and homogeneous $b$ and $|\chi|$, chiral phases with more complicated modulations occur.

The spatial patterns of the mean-field solutions are shown in detail in Fig.~\ref{fig:patt1}. While panel (a) shows the staggered TR-breaking flux pattern encountered in the FL$_c$ phase, panels (b) and (c) illustrate cases with more complicated modulations, one with TR-breaking fluxes (i.e. FL$_c^i$) and the other without TR-breaking fluxes (i.e. FL$_0^i$). We also note that, in some parameter regions, the details of the spatial modulations depend on the size of the real-space unit cell, i.e., the system wants to adopt a larger unit cell. Sample results for this fact, obtained for $N_{sc}=4\times4$, are in Fig.~\ref{fig:patt2}. We emphasize, however, that the checkerboard-flux FL$_c$ remains stable in large portions of the phase diagram also for unit cells larger than $N_{sc}=2\times 2$.


\begin{figure*}[!tb]
\includegraphics[width=0.93\textwidth]{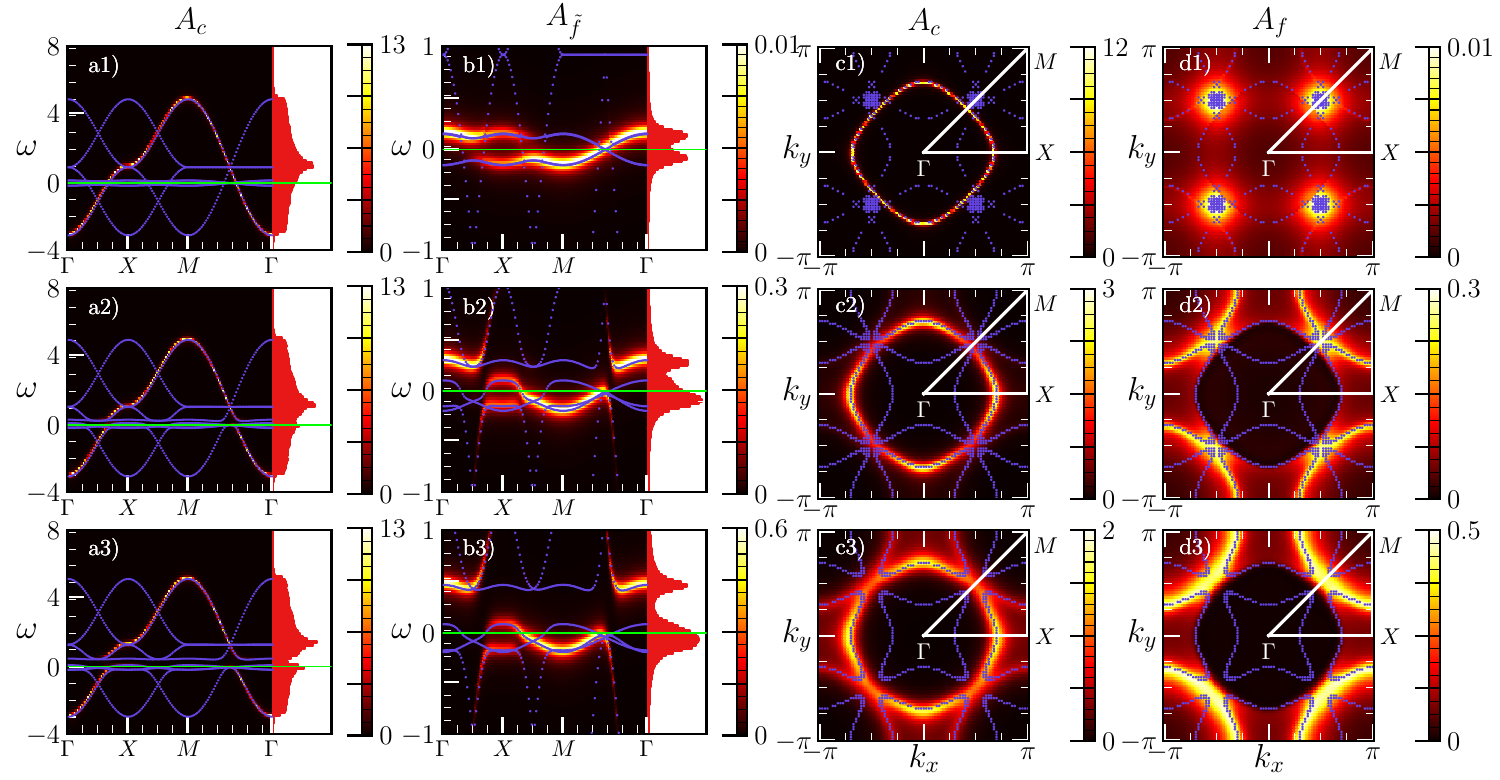}
\caption{
Single-particle spectral functions as in Fig.~3 of the main text, but now for $\nc=0.65$ and $\JK = 1.3$ (top), $1.7$ (middle), and $2.1$ (bottom), all located in the FL$_c$ phase with checkerboard flux pattern. Kondo screening gets stronger from top to bottom, for details see text.
}
\label{fig:spec2}
\end{figure*}

\subsection{Spectral function in the FL$_c$ phase}

Finally, we display a more detailed evolution of the single-particle spectrum across the chiral-metal phase FL$_c$ in Fig.~\ref{fig:spec2}. While all three parameter sets shown there have $b\neq 0$ and therefore correspond to a heavy Fermi liquid with a ``large'' Fermi surface, the quasiparticle band structure is complicated due to translation symmetry breaking arising from the checkerboard flux pattern. The Fermi-surface cut in panel (c) shows a gradual intensity transfer in $A_c$ from the small $c$-electron Fermi surface to the simple large Fermi surface of heavy quasiparticles (which would be present in the absence of translation symmetry breaking, see Fig.~3(c3,d3) of the main paper). This large Fermi surface is more prominent in $A_f$ shown in panel (d) except for the case of very small $b$ in panel (d1) where remnants of the spinon's Dirac band structure are visible.

We consider the band backfolding and the resulting disparity of the Fermi surfaces seen in $A_c$ and $A_{\tilde{f}}$ as a signature of the formation of the intermediate phase FL$_c$. High-resolution photoemission experiments with varying photon energy, resulting in varying matrix elements for $c$ and $f$ electrons, should be able to detect this.


\subsection{Circulating current in the FL$_c$ phase}

\begin{figure}[!b]
\includegraphics[width=\columnwidth]{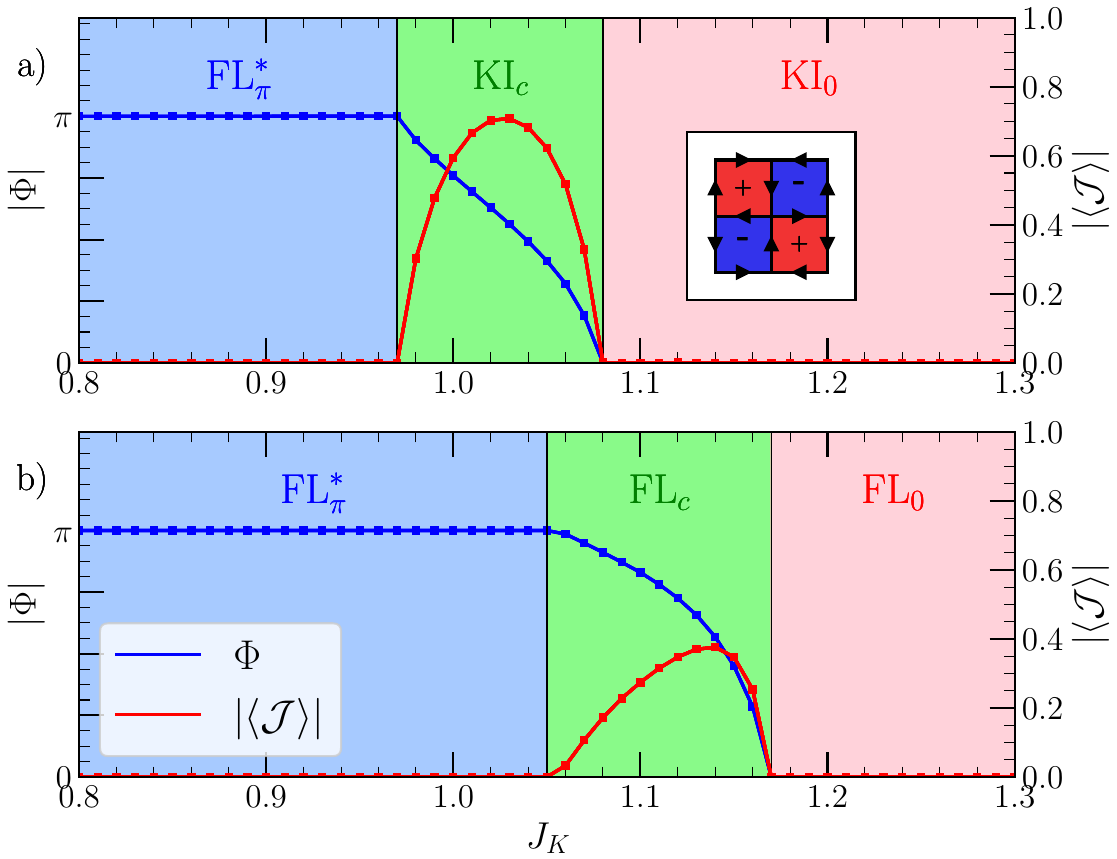}
\caption{
Bond current $\mathcal{J}$ and plaquette flux $\Phi$ for $T=0.05$, $\tilde{\kappa}=10$. Results are shown for (a) the chiral Kondo insulator $\nc=1$ and (b) the chiral metal $\nc=0.9$. The inset shows the spatial pattern of bond currents.
}
\label{fig:curr}
\end{figure}

The chiral metal FL$_c$ is an orbital antiferromagnet, with staggered circulating currents in the individual plaquettes. To demonstrate this, we compute the expectation value of the bond current operator
\begin{equation}
\mathcal{J}_{i \rightarrow j} = -\mathrm{i} t  (c_i^{\dagger}  c_j  -  c_j^{\dagger}  c_i )
\end{equation}
A sample set of numerical results is in Fig.~\ref{fig:curr}. TR-breaking fluxes $\Phi\neq 0,\pi$ are accompanied by non-zero equilibrium currents $\langle\mathcal{J}\rangle\neq 0$. These currents flow both in the FL$_c$ and KI$_c$ phases. Their magnitude is equal on all bonds, except in the  $\mathrm{FL}_c^i$ phases where additional modulations occur (not shown).

\section{Landau functional for the FL$_\pi^\ast$ -- FL$_c$ transition}

In the numerical calculations, we find a direct continuous transition from FL$_\pi^\ast$ to FL$_c$, with the simultaneous onset of Kondo screening ($b\neq0$) and TR-breaking fluxes ($\Phi\neq\pi$). Possible intermediate FL$_c^\ast$ and FL$_\pi$ phases are absent from the phase diagram. Hence, the transition from FL$_\pi^\ast$ to FL$_c$ constitutes a non-Landau transition even at the level of mean-field theory. Here we shed light onto this unusual behavior.

\begin{figure}[!tb]
\includegraphics[width=\columnwidth]{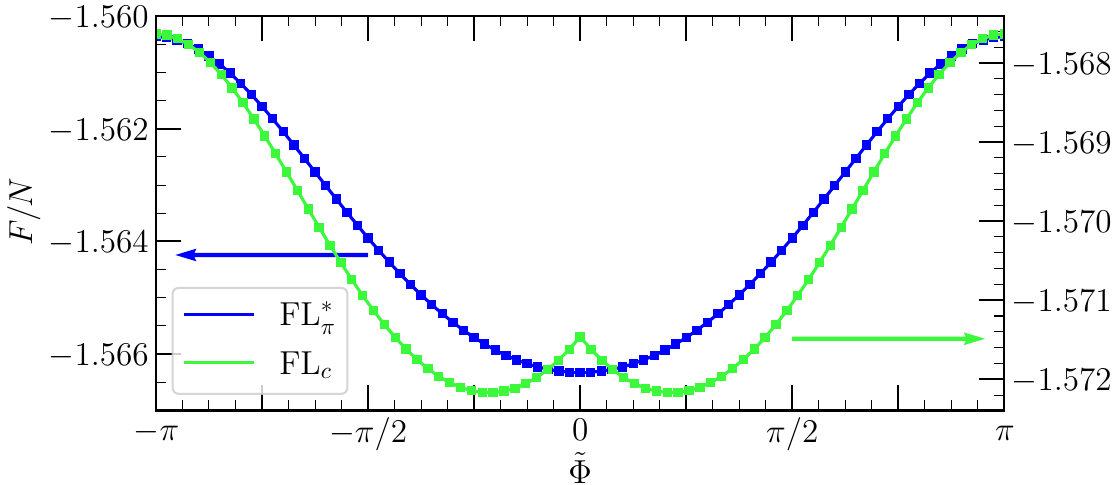}
\caption{
Free energy per site under variation of the emergent gauge flux for $\nc=0.6$ and $T=0.05$. $\tilde{\Phi} = \Phi-\pi$ is the deviation of the flux from the $\pi$-flux Dirac spin liquid. Data are shown for $J_{\mathrm{K}}=0$ ($\mathrm{FL}^{\ast}_{\pi}$) and $J_{\mathrm{K}}=1.8$ ($\mathrm{FL}_c$).
}
\label{fig:sing}
\end{figure}

To this end, we consider a Landau-type free energy, $F(\tilde\Phi)$ as function of the plaquette flux deviation from $\pi$, $\tilde\Phi=\Phi-\pi$, assuming a staggered flux pattern as in Fig.~\ref{fig:patt1} (a).
We can use our mean-field theory to explicitly compute $F(\tilde\Phi)$ by fixing the phases of the $\chi$ (and hence $\tilde{\Phi}$) but determining all other mean-field parameters self-consistently. Numerical results are in Fig.~\ref{fig:sing}. We see that $F(\tilde\Phi)$ displays a quadratic minimum at $\tilde\Phi=0$ in the FL$^\ast$ phase (i.e., for $b=0$) while it shows non-analytic behavior $\propto -|\tilde\Phi|$ in the FL$_c$ phase.
We conjecture a Landau expansion of the form
\begin{equation}
F(b, \tilde\Phi) = c_1 \tilde\Phi^2 + c_2 \tilde\Phi^4 + d_1 |b|^2 + d_2 |b|^4 - v |b|^2 |\tilde\Phi|
\label{landau}
\end{equation}
with coefficients $c_{1,2}$, $d_{1,2}$, and $v$ where $c_{1,2}>0$, $d_2>0$, and $d_1$ controlling the FL$_\pi^\ast$ -- FL$_c$ transition; higher-order terms have been omitted. As the mean-field spectrum is invariant under $\tilde{\Phi}\mapsto -\tilde{\Phi}$, $F$ must be an even function of $\tilde\Phi$. We also note that the complex phases of the $b$ are fixed up to gauge transformations once $\tilde\Phi$ is fixed, see below.

The coupling $v$ turns out to be positive, and it is the non-analyticity of this coupling term which is responsible for the non-Landau character of the transition. Indeed, assuming small positive $v$, minimizing the functional \eqref{landau} for $d_1>0$ yields
\begin{equation}
\tilde\Phi=0,~b=0
\end{equation}
while $d_1<0$ results in
\begin{align}
|\tilde\Phi|&=-\frac{1}{\sqrt[3]{3}}\frac{q}{p^{2/3}} \propto (-d_1), \nonumber\\
|b|&=\sqrt{\frac{v|\tilde{\Phi}| - d_1}{2d_2}} \propto \sqrt{-d_1}
\end{align}
with constants
\begin{equation}
p = \frac{c_1}{2c_2} - \frac{v^2}{8c_2d_2} > 0, q = \frac{d_1v}{8c_2d_2} < 0.
\end{equation}
We conclude that the Landau expansion \eqref{landau} describes the relevant features of the FL$_\pi^\ast$--FL$_c$ transition.

\begin{figure}[!b]
\includegraphics[width=\columnwidth]{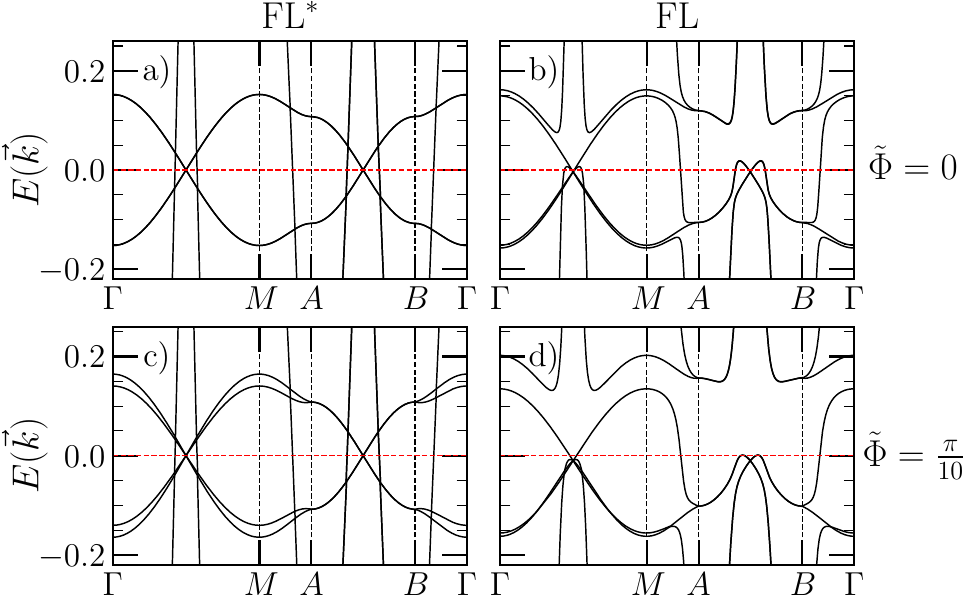}
\caption{
Mean-field band structures along the Brillouin-zone path in Fig.~\ref{fig:minipath} for fixed plaquette flux (a,b) $\tilde\Phi=0$ and (c,d) $\tilde\Phi=\pi/10$, in both the FL$^\ast$ phase (a,c, with $\JK=0$) and the FL phase (b,d, with $\JK=1.62$ and $1.70$, respectively), all with $\nc=0.6$. Thus, the panels correspond to phases (a) FL$_\pi^\ast$, (b) FL$_\pi$, (c) FL$_c^\ast$, and (d) FL$_c$.
}
\label{fig:bandcuts}
\end{figure}

\begin{figure}[!b]
\includegraphics[width=0.5\columnwidth]{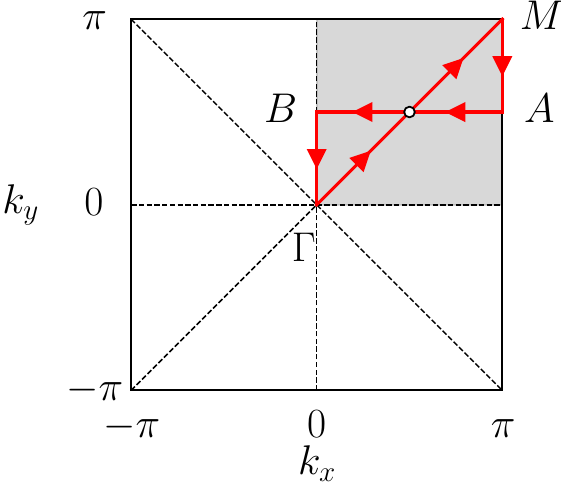}
\caption{
Brillouin-zone path employed in Fig.~\ref{fig:bandcuts}.
}
\label{fig:minipath}
\end{figure}

It remains to clarify the origin of the non-analytic coupling in Eq.~\eqref{landau}. We focus on $T=0$ where the mean-field energy is given by the zero-temperature limit of Eq.~\eqref{Free_Energy}; the generalization to finite $T$ is straightforward. Any non-analyticities must arise from features of the mean-field band structure. In the $f$-electron sector and for the $2\times2$ unit cell relevant for the checkerboard flux state, we have two degenerate Dirac cones at the Fermi level for $\tilde\Phi=0$, $b=0$, see Fig.~\ref{fig:bandcuts}(a). A non-zero $\tilde\Phi$ lifts this degeneracy and produces two anisotropic Dirac cones which are stretched in $\vec{k}=(\pi,\pi)^{\mathrm{T}}$ and $\vec{k}=(\pi,-\pi)^{\mathrm{T}}$ directions, respectively, Fig.~\ref{fig:bandcuts}(c). In fact, the $4\times 4$ Bloch matrix of the $f$-electron sector can be diagonalized analytically, and yields for the negative-energy bands
\begin{equation}
E_{1,2}(\vec{k}) = - 2|\tilde{\chi}||\vec{q}| \left\lbrace 1 \pm \frac{|\tilde{\Phi}|}{2} \hat{q}_x \hat{q}_y - \frac{\tilde{\Phi}^2}{8} \hat{q}_x^2\hat{q}_y^2 + \hdots \right\rbrace
\label{flstbands}
\end{equation}
where we have expanded in both $\tilde\Phi$ and the momentum deviation from the Dirac point, $\vec{q} = (\frac{\pi}{2},\frac{\pi}{2})^{\mathrm{T}} - \vec{k}$, and $\hat{q}_x,\hat{q}_y$ denote components of the corresponding unit vector. We also introduced $\tilde{\chi} := \tilde{J}_{\mathrm{H}}\chi(1-2\tilde{\kappa}|\chi|^2)$ for brevity. Summing over all negative-energy bands yields the fermionic contribution to the ground-state energy
\begin{align}
E = 4|\tilde{\chi}| \int \mathrm{d}^2 q \, |\vec{q}| &\left\lbrace -1 + \frac{\tilde{\Phi}^2}{8} \hat{q}_x^2\hat{q}_y^2
\right. \nonumber
\\
&\left.\quad + \frac{\tilde{\Phi}^4}{384} \hat{q}_x^2\hat{q}_y^2(15\hat{q}_x^2\hat{q}_y^2 + 4) + \hdots \right\rbrace \nonumber
\end{align}
which corresponds to the Landau form \eqref{landau} for $b=0$. We stress that the energy bands of FL$^{\ast}$ alone \eqref{flstbands} involve non-analytic contributions $\propto|\tilde{\Phi}|$, arising from the degeneracy of the Dirac cones present for $\tilde{\Phi}=0$, that only cancel out due to the integration over occupied bands.

\begin{figure}[!tb]
\includegraphics[width=\columnwidth]{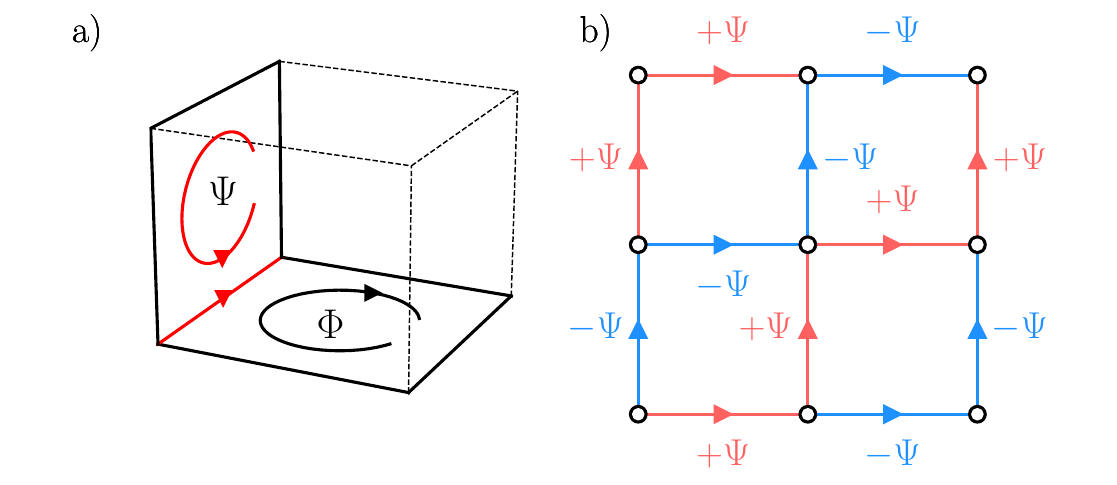}
\caption{
(a) Illustration of the Kondo plaquette fluxes $\Psi$ where the top (bottom) layer corresponds to $c$ ($f$) electrons, respectively.
(b) Sign convention for the Kondo fluxes, together with their preferred pattern in the FL$_c$ phase.
}
\label{fig:kondo_flux}
\end{figure}

Non-zero $b$, present in the FL phases, leads to additional hybridization with the $c$ band. At the mean-field level, the hoppings $f_i^{\dagger}f_j$ and $c_i^{\dagger}f_i$ allow one to define fluxes $\Psi$ in the ``Kondo plaquettes'', i.e., plaquettes formed by two $c$ and two $f$ sites, Fig.~\ref{fig:kondo_flux}(a). The fluxes are determined by the complex phases of $b$ and $\chi$, and one can easily show that the fluxes $\Psi$ in the Kondo plaquettes and the flux $\Phi$ in the $f$-electron plaquettes are related by
\begin{align}
\Phi_{\square} = \sum_{i\in\square} \Psi_i
\end{align}
where $\Psi_i$ correspond to the fluxes of the four Kondo plaquettes along one $f$ plaquette, properly oriented. The mean-field solution indicates a preferred Kondo flux pattern in FL$_c$, Fig.~\ref{fig:kondo_flux}(b), from which we conclude $\Phi\equiv -4\Psi\mod 2\pi$; this relates the phases of the $b$ to $\Phi$ up to gauge transformations.

We are now in the position to discuss the changes in the mean-field bands arising in FL from finite $b$, illustrated in Fig.~\ref{fig:bandcuts}. For $\tilde\Phi=0$ the strongly dispersive $c$ band hybridizes with both two Dirac bands, except along certain high-symmetry directions (like $\Gamma-M$) where one Dirac band remains untouched, Fig.~\ref{fig:bandcuts}(b). In addition, the Dirac point is shifted away from the Fermi level because $\nc\neq 1$ breaks particle--hole symmetry. Finite $\tilde\Phi$ now leads to an additional splitting of the (former) Dirac bands, Fig.~\ref{fig:bandcuts}(d). Since these Dirac bands are hybridized in an unequal fashion with the $c$ band, the $\tilde\Phi$-induced splitting affects them differently. Recalling that this splitting is $\propto|\tilde{\Phi}|$, this \textit{different} splitting prevents the cancellation of the $|\tilde{\Phi}|$ contributions in the momentum sum of the ground-state energy. The non-analytic term in Eq.~\eqref{landau} follows.

Let us summarize the ingredients for the non-analytic behavior: 
(i) The $\pi$-flux spin liquid displays half-filled spinon bands with Dirac cones.
(ii) Upon imposing a modulated flux pattern with fluxes different from $\pi$, the spinon bands split $\propto|\tilde{\Phi}|$. However, in the ground-state energy a contribution $\propto|\tilde{\Phi}|$ does not occur because the symmetry of the spinon band structure (which is higher than the lattice symmetry) leads to a cancellation. In the present square-lattice case, the $Z_4$ rotation symmetry about each Dirac point protects the analytic behavior.
(iii) The onset of Kondo screening leads to a hybridization with the $c$ band and breaks the protecting symmetry, i.e., reduces the symmetry of the spectrum to the lattice symmetries. As a result, a non-analytic term $\propto|b|^2|\tilde{\Phi}|$ occurs. In FL$_c$ with $n_c\neq 1$ particle--hole symmetry is broken, as in Fig.~\ref{fig:bandcuts}, but we stress that this is not essential, as the non-analytic term also arises in the Kondo insulator due to the broken $Z_4$ rotation symmetry about the Dirac point.

This argument strongly suggests that the simultaneous onset of Kondo screening and TR breaking is a generic feature of a Kondo-lattice system realizing a $\pi$-flux FL$^\ast$ phase.
